\documentclass[twocolumn,showpacs,preprintnumbers,amsmath,amssymb,superscriptaddress,pre,longbibliography]{revtex4-1}

\usepackage{graphicx}
\usepackage{dcolumn}
\usepackage{bm}
\usepackage{graphics}
\usepackage{subfigure}
\usepackage{graphicx}
\usepackage{epsfig}
\usepackage{epstopdf}
\usepackage{xcolor}
\usepackage{multirow}
\usepackage{mathtools}

\allowdisplaybreaks

\begin{document}

\title{Corrections to reaction-diffusion dynamics above the upper critical dimension}

\author{Johannes Hofmann}
\email{johannes.hofmann@physics.gu.se}
\affiliation{Department of Physics, Gothenburg University, 41296 Gothenburg, Sweden}

\date{\today}

\begin{abstract}
Reaction-diffusion models are common in many areas of statistical physics, where they describe the late-time dynamics of chemical reactions. Using a Bose gas representation, which  maps the real-time dynamics of the reactants to the imaginary-time evolution of an interacting Bose gas, we consider corrections to the late-time scaling of $k$-particle annihilation processes $k A \to \emptyset$ above the upper critical dimension, where mean-field theory sets the leading order. We establish that the leading corrections are not given by a small renormalization of the reaction rate due to $k$-particle memory effects, but instead set by higher-order correlation functions that capture memory effects of sub-clusters of reactants. Drawing on methods developed for ultracold quantum gases and nuclear physics, we compute these corrections exactly for various annihilation processes with $k>2$.
\end{abstract}

\maketitle

\section{Introduction}

Reaction-diffusion models describe the stochastic dynamics of particles that  spread diffusively and undergo local chemical reactions~\cite{hinrichsen00,krapivsky13,taeuber14}. They are ubiquitous in statistical physics, where they describe, for example, the dynamics of chemical reactions~\cite{kuzovkov88}, predator-prey populations~\cite{lotka10,volterra28}, or pattern formation~\cite{kondo10}. In particular, the specific case of $k$-particle annihilation
\begin{align}
k A \stackrel{\lambda}{\longrightarrow} \emptyset  \label{eq:annihilation}
\end{align}
 with a reaction rate~$\lambda$ describes processes such as the recombination of excitons in semiconductors~\cite{allam13}, monopole annihilation in models of the early universe~\cite{toussaint83}, reactions in polymer melts~\cite{deGennes82,deGennes82b}, or the dynamics of domain walls~\cite{family91}. Historically, this model was first investigated in a statistical physics context by von Smoluchowski to describe the coagulation kinetics in colloidal gold suspensions~\cite{smoluchowski16,smoluchowski17}. Of interest for annihilation processes like Eq.~\eqref{eq:annihilation} is the late-time dynamics of the reactant density $n(t)$ that characterizes the decay to the empty state~\cite{toussaint83,kang84,peliti86,meakin84},
which is independent of the initial reactant distribution. However, it depends sensitively on the space dimension $d$, since above an upper critical dimension $d_c = 2/(k-1)$~\cite{kang84,lee94} reactant particles are not correlated (at least to leading order), whereas below that they are~\cite{deGennes82}. The first case $d>d_c$ defines the reaction-limited regime, where the density (to a first approximation) solves a mean-field rate equation~\cite{taeuber14,krapivsky13}
\begin{align}
\partial_t n(t) = - k \lambda \, n^k(t) , \label{eq:meanfieldequation}
\end{align} 
which predicts a power-law decay at late times, 
\begin{align}
\lim_{t \to \infty} n_0(t) = \frac{1}{(k (k-1) \lambda t)^{1/(k-1)}} , \label{eq:meanfield} 
\end{align}
independent of the initial density but with an explicit dependence on the annihilation rate $\lambda$~\cite{cardy99,taeuber14}. The second case \mbox{$d<d_c$} defines the diffusion-limited regime where reactants are strongly anticorrelated, giving rise to a scaling \mbox{$n(t) \sim (Dt)^{-d/2}$} that is slower than the mean-field decay (here, $D$ is the diffusion constant), with $n(t) \sim [(\ln t)/Dt]^{1/(k-1)}$ at the critical dimension. This scaling, which is independent of the reaction rate $\lambda$, is called universal. Experimentally, diffusion-limited scaling has been observed in exciton recombination in semiconductors~\cite{kopelman86,prasad87,kroon93,monson03,russo06,allam13}. Theoretical work predominantly considers the diffusion-limited regime, which for integer dimensions describes the case $(k,d)=(2,1)$ as well as $(2,2)$ and $(3,1)$ with marginal scaling, using renormalization group methods~\cite{toussaint83,peliti86,droz93,lee94,howard96}, mappings to integrable models in one dimension~\cite{racz85,family91,alcaraz94,stinchcombe01}, and numerical simulations~\cite{meakin84,kang84,kang85,doering88,carlon99,carlon01}. By comparison, scaling in the reaction-limited regime appears less explored beyond the mean-field equation~\eqref{eq:meanfieldequation}, even though it describes most parameter combinations.

\begin{figure}[t!]
\scalebox{0.8}{\includegraphics{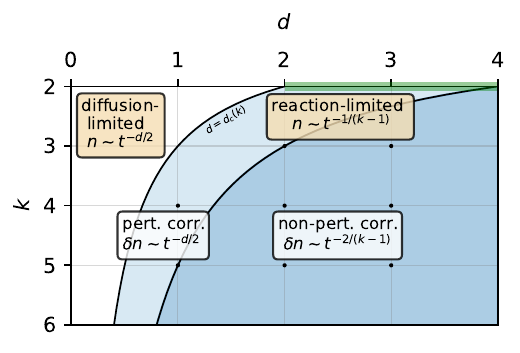}}
\caption{Asymptotic scaling behavior of reaction-diffusion processes with local $k$-particle annihilation as a function of the space dimension $d$. The continuous black line \mbox{$d=d_c=2/(k-1)$} marks the boundary between the diffusion-limited regime with \mbox{$n(t) \sim t^{-d/2}$} and the reaction-limited regime with mean-field scaling \mbox{$n(t) \sim t^{-1/(k-1)}$}. Corrections to mean-field scaling are set by memory effects of sub-clusters of the $k$ reactants, and they are either perturbative with \mbox{$\delta n(t) \sim t^{-d/2}$} (light blue shaded area) or nonperturbative with \mbox{$\delta n(t) \sim t^{-2/(k-1)}$} (dark blue shaded area), separated by the line \mbox{$d=4/(k-1)$} (black line). Moreover, the parameter case $(k,d) = (3,2)$ contains a logarithmic scaling correction. All corrections dominate over a simple renormalization of the reaction rate, except in the perturbative regime for $k=2$ (green line), where they are of the same order.}
\label{fig:1}
\end{figure}

The aim of this paper is to derive the corrections to mean-field scaling~\eqref{eq:meanfield} above the upper critical dimension \mbox{$d > d_c$}. By the argument given above, one could assume that this correction is set by a perturbative renormalization of the reaction rate $\lambda$ that corrects for reactant correlations. In detail, such a perturbation describes a memory effect that accounts for a reduction in the reactant density if the $k$ reactants have already met at some point in the past and annihilated. We show here that this is not correct. Instead, the leading-order scaling corrections are set by memory effects that account for a reactant depletion due to sub-clusters of $l < k$ reactants having reacted in the past with other particles, which are processes that involve a total particle number larger than~$k$. A quantitative discussion reveals two separate scaling regimes, which are summarized in Fig.~\ref{fig:1}: right above the critical dimension, the corrections are perturbative and describe a single past memory event, which leads to a scaling $\delta n(t) \sim t^{-d/2}$, whereas, for even higher dimensions, such terms must be summed to all orders, which gives a nonperturbative correction $\delta n(t) \sim t^{-2/(k-1)}$. In both regimes, the corrections are of higher order than the renormalization of the annihilation rate (at least for $k>2$). Corrections to mean-field scaling are thus more pronounced than one might expect. In addition, the magnitude of the corrections is parametrized by $\lambda$, which also parametrizes the non-universality of the leading term~\eqref{eq:meanfield}. The results of this paper should be observable in numerical simulations~\cite{vanzon05,opplestrup06}.
 
In deriving the scaling corrections, we make use of a representation of the reaction-diffusion system in terms of a bosonic Doi-Peliti path integral, which maps the process~\eqref{eq:annihilation} to a nonrelativistic Bose gas dual with non-Hermitian $k$-particle contact interactions. In this description, the diffusion constant $D$ corresponds to an inverse mass and the reaction rate $\lambda$ sets the strength of the interaction between bosons. Related (but not identical) models are used as effective field theories in atomic and nuclear physics, where they describe quantum gases of bosonic atoms or ${}^4$He~\cite{andersen04,braaten08}, as well as the scattering of neutrons or mesons~\cite{braaten06}. In particular, higher-order processes that determine the leading-order corrections to mean-field scaling are linked to vertex functions that describe the scattering of more than $k$ particles, and techniques to compute the three-body scattering amplitude in Bose quantum gases~\cite{braaten02,bedaque99a,bedaque99b} are applied to the problem.

The paper is structured as follows. We begin in Sec.~\ref{sec:doipeliti} with a discussion of the Doi-Peliti path integral. Next, in Sec.~\ref{sec:effectiveaction}, we derive a dynamical equation for the density using the effective action, which systematically includes beyond-mean-field corrections through the vertex functions. We establish a power counting for these vertex functions and show that the leading-order corrections to mean-field scaling stem from higher-order vertices. To obtain a result that is independent of a short-distance cutoff, some vertex functions must be summed to all orders, which is done numerically for various decay processes and dimensions. Section~\ref{sec:summary} contains a summary and outlook.

\section{Doi-Peliti path integral}\label{sec:doipeliti}

We begin by introducing the representation of the reaction-diffusion system~\eqref{eq:annihilation} in terms of a bosonic Doi-Peliti path integral~\cite{doi76,grassberger80,peliti85}. Reviews of the Doi-Peliti formalism and reaction-diffusion systems are found in Refs.~\cite{cardy96,mattis98,cardy99,taeuber05,taeuber14} and of the effective field-theory description of Bose quantum gases in Refs.~\cite{braaten06,braaten08}.

To capture the dynamics of the process~\eqref{eq:annihilation} beyond a mean-field approximation, consider first a microscopic model for the reaction-diffusion system~\cite{taeuber05} defined on a lattice with lattice constant $a_0$. A lattice site with index $i$ is occupied by $n_i$ particles, and if this number is larger than $k$, particles can annihilate according to the prescription~\eqref{eq:annihilation} with a bare annihilation rate $g_0$. Ultimately, we are interested in aspects of the model that do not depend on the lattice spacing $a_0$, i.e., we will take the continuum limit.

Denote the occupation probability for $\{n_i\}$ particles on the lattice sites by $P(\{n_i\}; t)$. It evolves in time according to the master equation
\begin{align}
\frac{\partial P(\{n_i\}; t)}{\partial t} &=
g_0 \sum_i \biggl\{\frac{(n_i + k)!}{n_i!} P(\ldots, n_i+k, \ldots; t) \nonumber \\
&\quad - \frac{n_i!}{(n_i-k)!} P(\ldots, n_i, \ldots; t)\biggr\} , \label{eq:master}
\end{align}
with additional terms that account for hopping, i.e., diffusion, between lattice sites. Here, the first term describes a gain as $k$ particles annihilate at a site $i$ with $n_i + k$ particles and the second term describes a loss as $k$ particles are removed from a state with $n_i$ particles. 

To recast this equation in a Fock-space formalism, define the ket vector $|n_i\rangle$ that denotes a single-site state with $n_i$ particles. We introduce bosonic creation and annihilation operators $a_i^\dagger$ and $a_i$, which act on a single-site state as
\begin{align}
a_i |n_i\rangle &= n_i |n_i - 1\rangle \\
a_i^\dagger |n_i\rangle &= |n_i + 1\rangle .
\end{align}
This convention is different from the usual bosonic ladder operators in quantum mechanics~\cite{sakurai94}, but the number operator acts in the same way as $a_i^\dagger a_i |n_i\rangle = n_i |n_i\rangle$. In particular, $|n_i\rangle = (a_i^\dagger)^{n_i} |0\rangle$ with $|0\rangle$ the single-site vacuum state. The Hilbert space of the full lattice is spanned by the direct product of single-site Hilbert spaces. A state with definite particle number on each lattice site is then represented by the Fock state
\begin{align}
|\{n_i\}\rangle &= \prod_i (a_i^\dagger)^{n_i} |0\rangle , \label{eq:deffock}
\end{align}
where $|0\rangle$ is the many-site vacuum state. Now, define the state vector
\begin{align}
|\Psi(t)\rangle = \sum_{\{n_j\}} P(\{n_j\}; t) |\{n_j\}\rangle , \label{eq:statevector}
\end{align}
which, from Eq.~\eqref{eq:master}, obeys an imaginary-time Schr\"odinger equation 
\begin{align}
\partial_t |\Psi(t)\rangle = - H |\Psi(t)\rangle
\end{align}
with a non-Hermitian Hamiltonian that does not involve combinatorial factors~\cite{taeuber05,taeuber14}:
\begin{align}
H = g_0 \sum_i [1-(a_i^\dagger)^k] a_i^k .
\end{align}
Here, the first term represents the gain term in Eq.~\eqref{eq:master} and the second term the loss term. 
The state $|\Psi(t)\rangle$ then evolves as $|\Psi(t)\rangle = e^{- H t} |\Psi_0\rangle$ with an initial state $|\Psi_0\rangle$. Likewise, an additional hopping term with bare hopping amplitude $D_0$ between nearest neighbor sites $\langle ij \rangle$ is represented by a term \mbox{$D_0 \sum_{\langle ij \rangle} (a_i^\dagger - a_j^\dagger) (a_i - a_j)$}~\cite{taeuber05}.

The average particle number at a lattice point $r$ is expressed in terms of the state vector as~\cite{cardy96,taeuber05}
\begin{align}
\langle N(t) \rangle = \sum_{\{n_j\}} n_r \, P(\{n_j\}; t) = \langle {\cal P} | a_r e^{- H t} |\Psi_0\rangle , \label{eq:expectation}
\end{align}
where $\langle {\cal P} | = \langle 0 | \prod_i e^{a_i}$ is a coherent projection state. The first equality is the definition of the expectation value and the second equality follows from \mbox{$\langle {\cal P} | 0 \rangle = 1$} and \mbox{$\langle {\cal P} | a_i^\dagger = \langle {\cal P} |$}. Note that the form~\eqref{eq:expectation} differs from the quantum-mechanical definition of the expectation value, which involves the square of the wave vector. Equation~\eqref{eq:expectation} can be expressed as a coherent-state path integral with coherent states $|\phi\rangle = e^{\sum_i \phi_i a_i^\dagger} |0\rangle$, where the $\phi_i$s are the eigenvalues at site $i$, i.e., $a_i |\phi\rangle = \phi_i |\phi\rangle$. Formally, such a state describes (up to normalization) a Poisson distribution of $n_i$-particle states at site $i$. This gives
\begin{align}
n(t) &= \langle \phi(t) \rangle = \int {\cal D}[\bar{\phi}, \phi] \, \phi(t) e^{- {\cal A}[\bar{\phi}, \phi]} , \label{eq:density}
\end{align}
where the path-integral measure is defined as \mbox{${\cal D}[\bar{\phi}, \phi] = \prod_i d\phi_i^* d\phi_i/(2\pi i)$} and the term ${\cal A}$ in the exponent is known as the Doi-Peliti action. Taking the continuum limit with a coupling \mbox{$g = a_0^{(k-1)d} g_0$} and diffusion constant \mbox{$D=a_0^2 D_0$}, it reads (neglecting boundary terms)
\begin{align}
&{\cal A}[\bar{\phi}, \phi] \nonumber \\
&= 
\int d^dx \, \int_0^t dt' \, \, \biggl[ \bar{\phi} \biggl(\frac{\partial \phi}{\partial t'} - D \nabla^2 \phi\biggr)
- g (1-\bar{\phi}^k) \phi^k
\biggl] . \label{eq:doipeliti}
\end{align}
Here, $\phi$ is a bosonic field of length dimension $-d$ whereas $\bar{\phi}$ is dimensionless. If one identifies the diffusion constant with an inverse mass, $D=\hbar^2/2m$, the Doi-Peliti action~\eqref{eq:doipeliti} is similar (but not identical) to the effective description of a dilute Bose quantum gas, for which the term $g \bar{\phi}^k \phi^k$ in Eq.~\eqref{eq:doipeliti} describes the scattering of $k$ bosons via a contact interaction. The theories differ in the non-Hermitian vertex $- g \phi^k$ that would describe the annihilation of $k$ bosons~\footnote{
Note, however, that ultracold Bose quantum gases are metastable and decay through three-body recombination~\cite{esry99,bedaque00,braaten06,dincao18}, which is described by a reaction-diffusion process with \mbox{$k=3$}. Indeed, experimentally the three-body decay is accurately captured to leading order by the mean-field expression~\eqref{eq:meanfieldequation} with $k=3$~\cite{rem13,fletcher13,makotyn14,eismann16,eigen17}, where the rate $\lambda$ can acquire a microscopic dependence on the density, which may change the scaling exponents. A special case is the so-called unitary limit with \mbox{$\lambda \sim n^{-4/3}$}, which (at this level) indicates a scale-invariant decay without scaling corrections.
}.

\begin{figure}[t]
\begin{alignat}{4}
\raisebox{-0.03cm}{\scalebox{0.4}{\includegraphics{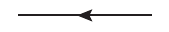}}} &= \tfrac{1}{s + D q^2}\quad &
\raisebox{-0.08cm}{\scalebox{0.6}{\includegraphics{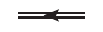}}} &= - \frac{1}{g} & 
\raisebox{-0.45cm}{\scalebox{0.4}{\includegraphics{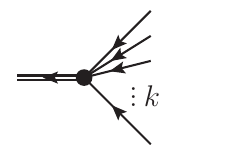}}} \hspace{-0.5cm} &= - g
 \nonumber \\[2ex]
\raisebox{-0.45cm}{\scalebox{0.4}{\includegraphics{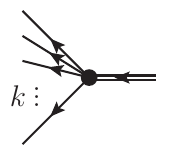}}} &= - g \qquad &
\raisebox{-0.45cm}{\scalebox{0.4}{\includegraphics{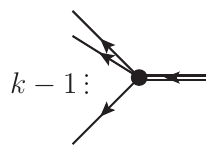}}} \hspace{-0.0cm} &= - k g \quad \ldots \, &
\raisebox{-0.07cm}{\scalebox{0.45}{\includegraphics{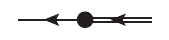}}} \hspace{-0.0cm} &= - k g & \nonumber
\end{alignat}
\caption{Feynman rules for the interaction vertices of the Doi-shifted action~\eqref{eq:doipeliti3}. Continuous single lines denote the propagator of the $\phi$ field and double lines the auxiliary $k$-particle field $d$.}
\label{fig:2}
\end{figure}

For further calculations, it is convenient to rewrite the Doi-Peliti action with a nondynamical auxiliary $k$-particle field \mbox{$d = \phi^k$}~\cite{braaten06} through a Hubbard-Stratonovich transformation of Eq.~\eqref{eq:doipeliti}:
\begin{align}
&{\cal A}[\bar{\phi}, \phi, \bar{d}, d] = \int d^dx \, \int_0^t dt' \, \biggl[\bar{\phi} \biggl(\frac{\partial \phi}{\partial t'} - D \nabla^2 \phi\biggr) \nonumber \\*
&\qquad \qquad\qquad \, - g (\bar{d} - \bar{\phi}^{k}) d - g (1 - \bar{d}) \phi^k 
\biggr] . \label{eq:doipeliti2}
\end{align}
This is a common representation in nonrelativistic field theories~\cite{braaten06} that simplifies diagrammatic calculations considerably. In addition, since the field operators are not normal ordered with respect to the projection state $\langle {\cal P} |$, it is customary to perform a ``Doi shift'' of the conjugate fields in the Doi-Peliti action~\eqref{eq:doipeliti2} as  $\bar{\phi} \to 1 + \bar{\phi}$ and $\bar{d} \to 1 + \bar{d}$~\cite{cardy96}:
\begin{align}
&{\cal A}'[\bar{\phi}, \phi, \bar{d}, d] = \int d^dx \, \int_0^t dt' \, \biggl[\bar{\phi} \biggl(\frac{\partial \phi}{\partial t'} - D \nabla^2 \phi\biggr) \nonumber \\*
&\qquad \qquad\qquad \, - g \bar{d} d + g \sum_{i=1}^k \binom{k}{i} \bar{\phi}^{i} d + g \bar{d}\phi^k
\biggr] . \label{eq:doipeliti3}
\end{align}
 Note that Eq.~\eqref{eq:density} can now be written as
 \begin{align}
 n(t) = \frac{\delta {\cal Z}}{\delta j} \Bigr|_{j,\bar{j}=0} ,
 \end{align}
 with a generating functional
 \begin{align}
 {\cal Z}[j,\bar{j}] = \int {\cal D}[\bar{\phi}, \phi, \bar{d}, d] \, e^{- {\cal A}'[\bar{\phi}, \phi, \bar{d}, d] + \int_{t,{\bf r}} (\bar{j} \bar{\phi} + j \phi)} \label{eq:Z}
 \end{align}
that contains source fields $j$ and $\bar{j}$. Feynman rules for this theory are as follows (adhering to the convention of Ref.~\cite{taeuber14},  which avoids symmetry factors in the action): imaginary time runs from the right to the left in a Feynman diagram. Continuous lines represent single-particle propagators, which carry a momentum label ${\bf q}$ and contribute a factor $G_0(t, {\bf q}) = \Theta(t) e^{-Dq^2t}$, and double lines the nondynamical field $d$, which contributes $-\delta(t)/g$. Feynman rules are shown in Fig.~\ref{fig:2}, where we state the Laplace transform of propagators and vertices defined as $f(s) = \int_0^\infty dt \, e^{-st} f(t)$, which depends on a frequency variable $s$, with the inverse Laplace transform $f(t) = \int_{\rm BW} \frac{ds}{2\pi i} \, e^{st} f(s)$, where ${\rm BW}$ is the Bromwich contour. Due to the Doi shift, there is only one vertex that describes the fusion of $k$ bosons to a $k$-boson line, but several that describe the splitting of the line into \mbox{$l = 1,\ldots,k-1$} particles, with corresponding Feynman rule $- g \binom{k}{l}$. Note that this does not imply that less than $k$ reactant particles annihilate. Diagrams carry a combinatorial factor that accounts for the multiplicity of vertices and different ways of connecting the propagator lines, and vertex functions have an overall minus sign. Momentum conservation is imposed at every vertex and undetermined loop momenta and time labels are integrated over.

\section{Effective action}\label{sec:effectiveaction}

The Doi-Peliti generating functional~\eqref{eq:Z} is linked to an equation of motion for the density through the effective action, which systematically takes into account fluctuations. In this section, we work out the corrections to mean-field scaling using this formalism. We begin in Sec.~\ref{sec:effactionA} by reproducing the mean-field result~\eqref{eq:meanfield} and derive a first correction due to a $k$-particle memory effect (discussed already in the Introduction), which however is not of leading order. As illustrated in Fig.~\ref{fig:1}, there are instead two distinct regions with different leading-order corrections: a perturbative correction, which involves a two-particle memory correction and which is derived in Sec.~\ref{sec:effectiveactionB}, and a nonperturbative correction, which involves a \mbox{$(k-1)$}-particle memory correction and which is derived in Sec.~\ref{sec:effectiveactionC}.

The effective action is defined in terms of the generating functional ${\cal Z}[j, \bar{j}]$ by a Legendre transformation with respect to the field expectation values \mbox{$\Phi = \langle \phi\rangle$} and \mbox{$\bar{\Phi} = \langle \bar{\phi} \rangle$}~\cite{amit05,kamenev11,taeuber14}:
\begin{align}
\Gamma[\bar{\Phi},\Phi] &= - \ln {\cal Z}[j, \bar{j}] + \int d^dx \int_0^t dt' \, (\bar{j} \bar{\Phi} + j \Phi) . 
\end{align}
It may be expanded in powers of $\Phi$ and $\bar{\Phi}$ with coefficients set by the vertex functions
\begin{align}
&\bar{\Gamma}_{\bar{N},N}(\bar{t}_1,\ldots; t_1, \ldots) \nonumber \\*
&\qquad= \frac{\delta \Gamma[\bar{\Phi},\Phi]}{\delta \bar{\Phi}(t_1) \ldots \delta \bar{\Phi}(t_{\bar{N}}) \delta \Phi(t_1) \ldots \delta \Phi(t_N)} \biggr|_{\Phi,\bar{\Phi}=0} , \label{eq:vertexfunction}
\end{align}
where we assume homogeneous field configurations in the following. Diagrammatically, the vertex functions $\bar{\Gamma}_{\bar{N},N}$ describe one-particle irreducible (1PI)  processes with $N$ ingoing and $\bar{N}$ outgoing lines at zero momentum. In terms of the Bose gas representation, they represent the 1PI scattering of an initial state with $N$ bosons to a final state with $\bar{N}$ bosons. The standard identities $\delta \Gamma/\delta \Phi = j$ and $\delta \Gamma/\delta \bar{\Phi} = \bar{j}$ then define an equation of motion for the fields $\bar{\Phi}$ and $\Phi$ by varying $\Gamma[\bar{\Phi}, \Phi]$ in the absence of sources. The first variation with respect to $\Phi$ gives $\bar{\Phi} = 0$, which is required by probability conservation~\cite{cardy96}, and the variation with respect to $\bar{\Phi}$ gives an equation of motion for the density $n = \Phi |_{j,\bar{j}=0}$:
\begin{align}
\frac{\delta \Gamma}{\delta \bar{\Phi}(t)} \biggr|_{\bar{\Phi},j,\bar{j}=0} &= 0 . \label{eq:eom}
\end{align}
Note that only vertices with a single outgoing line (\mbox{$\bar{N} = 1$}) will contribute to the dynamical equation. In the following, we use the notation $\Gamma_l$ for the vertex $\Gamma_{1,l}$, where the missing bar indicates that we separate all delta functions in time.

\begin{figure}[t!]
\begin{align}
\Gamma_1 &= \raisebox{-0.03cm}{\scalebox{0.4}{\includegraphics{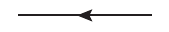}}} \nonumber \\[2ex]
\Gamma_k &= \raisebox{-0.3cm}{\scalebox{0.35}{\includegraphics{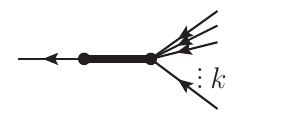}}} \hspace{-0.3cm} \nonumber \\[1ex]
&\hspace{-0.8cm}\scalebox{0.5}{\includegraphics{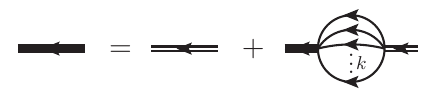}}\nonumber
 \end{align}
\caption{First two vertex functions that contribute to the equation of motion~\eqref{eq:eom}. They capture both the scaling crossover for $d \leq d_c$ and the mean-field result for $d>d_c$.}
\label{fig:3}
\end{figure}

\subsection{Mean-field solution}\label{sec:effactionA}

The leading-order terms in the equation of motion~\eqref{eq:eom} that involve the smallest power of the density are set by the vertices $\Gamma_1 = -G_0^{-1}$ and $\Gamma_k$, which are shown in Fig.~\ref{fig:3}. The corresponding equation for the density reads
\begin{align}
\partial_t n &= \int_0^t dt' \, \Gamma_k(t-t') n^k(t') , \label{eq:mfeom}
\end{align}
where we omit a boundary term $\bar{n} \delta(t)$ that sets the initial density $\bar{n}$. In defining the vertex $\Gamma_k$, we separate a $k$-particle propagator that is indicated by a bold line in Fig.~\ref{fig:3}. The equation for $\Gamma_k$ is
\begin{align}
\Gamma_k(t) &= - k g \delta(t) - g k! \int_0^t dt' \, \Gamma_k(t-t') S_k(t') , \label{eq:bethesalpeter}
\end{align}
where $k!$ is a symmetry factor for the different ways of combining the $k$ boson lines in the loop integral [note that our definition of the vertex functions~\eqref{eq:vertexfunction} implies that there is no symmetry factor associated with the ordering of the $k$ ingoing lines]. Furthermore, we define the loop integral 
\begin{align}
S_k(t) &= \int_{{\bf p}_1} \ldots \int_{{\bf p}_k} \, \delta({\bf p}_1 + \ldots + {\bf p}_k) \prod_{i=1}^k G(t,{\bf p}_i) ,
\end{align}
which is called the memory function and which is the diffusion propagator of $k$ bosons from one identical point in space to another identical point. Intuitively, the first term in Eq.~\eqref{eq:bethesalpeter} describes the reaction rate given an uncorrelated reactant distribution. The convolution integral then describes a memory effect that accounts for anticorrelations due to processes where $k$ particles have already reacted in the past~\cite{deGennes82}. Following the discussion in the Introduction, we expect that the first (second) term in Eq.~\eqref{eq:bethesalpeter} dominates above (below) the critical dimension. Indeed, Eq.~\eqref{eq:mfeom} is solved using a Laplace transformation~\cite{deGennes82,lee94}
\begin{align}
s n(s) &= \Gamma_k(s) [n^k](s) , \label{eq:mfeomlaplace}
\end{align}
where we denote by $[n^k](s)$ the Laplace transform of $n^k(t)$. Using the convolution theorem, Eq.~\eqref{eq:bethesalpeter} forms a geometric series that evaluates to
\begin{align}
\Gamma_k(s) &= - k \Bigl[\frac{1}{g} + k! S_k(s)\Bigr]^{-1}  \label{eq:gammas}
\end{align}
with the Laplace transform of the memory function
\begin{align}
S_k(s) &= \begin{cases}
\frac{\Gamma(1-\frac{d}{d_c})}{D k^{d/2} (4\pi)^{d/d_c}} \Bigl(\dfrac{D}{s}\Bigr)^{1-\frac{d}{d_c}}  & d<d_c(k) \\[3ex]
- \frac{1}{4\pi D k^{d_c/2}} \ln \dfrac{s}{D \Lambda^2}   & d=d_c(k) \\[2ex]
\frac{1}{D k^{d/2} (4\pi)^{d/d_c(k)} \Gamma(\frac{d}{d_c})} \dfrac{\Lambda^{2 (\frac{d}{d_c}-1)}}{d/d_c-1}   & d>d_c(k) \\[2ex]
\quad + \frac{\Gamma(1-\frac{d}{d_c})}{D k^{d/2} (4\pi)^{d/d_c}} \Bigl(\dfrac{s}{D}\Bigr)^{\frac{d}{d_c}-1} . &
\end{cases}  \label{eq:Gl}
\end{align}
Here, $\Lambda$ is a momentum space cutoff and we recall that $d_c = 2/(k-1)$ is a function of $k$. The expression is finite for \mbox{$d<d_c$}, and there is a logarithmic divergence for \mbox{$d=d_c$} and a power-law divergence for \mbox{$d>d_c$}. This strong dependence on a short-distance scale $r_0 \simeq \Lambda^{-1}$ indicates that the contact potential is not a well-defined reaction potential for \mbox{$d\geq d_c$}. For \mbox{$k=2$} (\mbox{$d_c=2$}), this is linked to the lack of re-entrance for Brownian motion in higher dimensions~\cite{taeuber05}, such that two point particles starting at different positions will never meet and thus never react unless the reaction potential has a more complicated short-distance form with a finite range~$r_0$~\cite{smoluchowski16,smoluchowski17}. However, as pointed out by~\textcite{deGennes82}, at time and distance scales that are much larger than $r_0^2/D$ and $r_0$, the annihilation vertex is still of the form~\eqref{eq:gammas} with an effective rate $\lambda$. Formally, for $d>d_c$, the UV-divergence in the memory function in Eq.~\eqref{eq:Gl} may be absorbed into a redefinition of the rate~$g$
\begin{align}
\frac{1}{\lambda} &= \frac{1}{g} + \frac{k!}{D k^{d/2} (4\pi)^{d/d_c} \Gamma(d/d_c)} \frac{\Lambda^{2 (\frac{d}{d_c}-1)}}{d/d_c-1} .
\end{align}
The effective rate $\lambda$ defines a characteristic length scale $b$ via $\lambda/D \sim b^{2 (d-d_c)/d_c}$, which is called the capture radius and which is (in principle) independent of $r_0$~\cite{deGennes82}. For \mbox{$d=d_c$}, where $g$ is dimensionless, the bare coupling is linked to a capture radius by dimensional transmutation as $b = \Lambda^{-1} \exp[-2Dk^{d_c/2}\pi/gk!]$, which is known as a scale anomaly~\cite{pitaevskii97,olshanii10,hofmann12}. The renormalized vertex is
\begin{align}
\Gamma_k(s) &= \begin{cases}
\frac{k}{4\pi D k^{d_c/2}} \ln \dfrac{s b^2}{D} & d=d_c \\[1ex]
 - k \Bigl[\frac{1}{\lambda} + \frac{k! \Gamma(1-\frac{d}{d_c})}{D k^{d/2} (4\pi)^{d/d_c}} \Bigl(\dfrac{s}{D}\Bigr)^{\frac{d}{d_c}-1}\Bigr]^{-1} & d\neq d_c
 \end{cases}  \label{eq:gammas2}
\end{align}
and no longer contains a strong cutoff dependence. Equation~\eqref{eq:gammas2} is valid below $d < 2d_c$ and additional logarithmic divergences appear at integer multiples of $d_c$. They can be renormalized by including higher-order reaction terms that include derivatives, but they will not contribute to the vertex function in the limit $s\to 0$, which is the one relevant in this paper. 

Note that this discussion of the $k$-particle memory function in reaction-diffusion systems is similar to that of scattering in quantum gases, where Eq.~\eqref{eq:gammas} describes the scattering $T$ matrix of $k$ bosons (typically, $k=2$) via a contact interaction~\cite{braaten08}. The renormalization then links the strength of the contact interaction to the $s$-wave scattering length $\lambda/D \sim a^{2 (d-d_c)/d_c}$, which is the universal parameter that encodes all information about low-energy scattering via a (possibly unknown) short-range potential.

To solve Eq.~\eqref{eq:mfeomlaplace}, impose the power-law scaling \mbox{$n(t) = A t^{-\alpha}$} at late times, which implies \mbox{$n(s) = A \Gamma(1-\alpha) s^{\alpha-1}$} and \mbox{$[n^k](s) = A^k \Gamma(1-k\alpha) s^{k\alpha-1}$} at small $s$ [at \mbox{$d=d_c$}, use \mbox{$n(t) = A (\ln t/t)^{\alpha}$}]. Below \mbox{$d<d_c$}, the vertex interpolates between the mean-field expression \mbox{$\lim_{s\to\infty} \Gamma_k(s) = - k \lambda$} at large $s$ (small times) and the diffusion limit \mbox{$\lim_{s\to0} \Gamma_k(s) \sim - k s^{(d_c-d)/d_c}$} at small $s$ (late times). Thus, provided that \mbox{$(\lambda/D) \bar{n}^{(d_c-d)/d} \gg 1$}---i.e., if the initial density is negligible---the density scaling will transition from a reaction-limited mean-field decay with exponent \mbox{$n(t) \sim (\lambda t)^{-1/(k-1)}$} at early times to the (slower) diffusion-limited decay with \mbox{$n(t) \sim (Dt)^{-d/2}$} at late times. In the special case \mbox{$d=d_c$}, we find \mbox{$\Gamma_k(s\to0) \to k/(\ln sb^2/D)$} and \mbox{$n(t) \sim [(\ln t)/t]^{1/(k-1)}$}, i.e., the mean-field result with a logarithmic scaling correction. The scaling crossover below \mbox{$d<d_c$} from the reaction-limited to the diffusion-limited regime has been observed in exciton recombination in one-dimensional carbon nanotubes~\cite{allam13}. However, an analogous crossover for \mbox{$d>d_c$} does not exist. To leading order at late times (small $s$), we have \mbox{$\Gamma_k(s) = - k \lambda$}, which reproduces the mean-field result~\eqref{eq:meanfield}, but since the memory function~\eqref{eq:Gl} has negative sign, the vertex diverges as the scale is increased to $s \simeq Db^{-2}$. This is known as a Landau pole~\cite{throckmorton15}, which marks the limit of the description in terms of a contact interaction and is absent if a microscopic potential (such as hard-core potential) is used.

Nevertheless, Eq.~\eqref{eq:mfeomlaplace} still sets a correction to the mean-field scaling that is obtained by expanding the vertex~$\Gamma_k(s)$ to leading order in $\lambda$. Expanding around the mean-field result $n = n_0 + \delta n$, the perturbation solves
\begin{align}
s \delta n(s) &= - k^2 \lambda \, [n_0^{k-1} \delta n](s) - k \, \delta \lambda(s) [n_0^k](s) ,
\end{align}
with 
\begin{align}
\delta \lambda(s) = - \frac{\lambda^2 k! \Gamma(1-d/d_c)}{D k^{d/2} (4\pi)^{d/d_c}} \Bigl(\frac{s}{D}\Bigr)^{(d-d_c)/d_c} ,
\end{align}
where the external fields  in the subleading term of the vertex function are evaluated at the mean-field value. In real time, the solution is $\delta n(t) = B t^{-\beta}$ with an exponent $\beta = (k-1)^{-1} + (d-d_c)/d_c$. 

Note that this scaling also follows from dimensional analysis as the correction $\delta n$ is suppressed by ${\cal O}(\lambda)$ compared to the mean-field equation and must be a function of the small dimensionless parameter $\lambda/[D (Dt)^{(d-d_c)/d_c}] \ll 1$. In the next section, we establish that this $k$-particle memory term does not form the leading correction to mean-field scaling, but that there are higher-order vertex corrections that describe memory effects of sub-clusters of reactants.

\subsection{Perturbative scaling corrections}\label{sec:effectiveactionB}

It is straightforward to obtain higher-order vertices starting from any given vertex $\Gamma_m$ (such as the one in Fig.~\ref{fig:3}) by pinching a number of $l<k$ (where $k>2$) ingoing lines and fusing them to a $k$-particle line at an earlier time (using the vertices in the second line of Fig.~\ref{fig:2}), which generates a contribution to the vertex $\Gamma_{m+k-l}$. These higher-order (in the external fields) vertex functions account for memory effects that describe anticorrelations in the reactant density at a time \mbox{$t' < t$} due to a subcluster of $l$ particles having reacted in the past with $k-l$ other reactants. To determine the order of the vertex contributions to the dynamical equation at small $\lambda$ for $d>d_c$, we replace the bold $k$-particle line by its mean-field value $-\lambda$ and use the mean-field scaling ${\cal O}(\lambda^{-1/(k-1)})$ for the external fields. The contribution of this new vertex to the equation of motion is then suppressed by ${\cal O}(\lambda^{(l-1)/(k-1)}) < {\cal O}(\lambda)$ compared to the contribution of the original vertex. Note that this power counting assumes that the vertices are finite, which is not always the case and will be revisited in the next section. 

\begin{figure}[t!]
\begin{align}
\Gamma_{2k-2} &= \raisebox{-0.65cm}{
$\stackrel{
\raisebox{-0.3cm}{
\scalebox{0.35}{\includegraphics{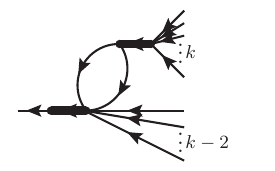}}
}}{\scriptscriptstyle {\cal O}(\lambda^{1/(k-1)})\quad} $
}\hspace{-0.3cm} 
 \nonumber \\
\Gamma_{2k-3} &= \raisebox{-0.55cm}{
$\stackrel{
\raisebox{-0.3cm}{
\scalebox{0.35}{\includegraphics{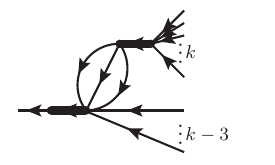}}
}}{\scriptscriptstyle {\cal O}(\lambda^{2/(k-1)})\quad} $
}\hspace{-0.3cm} \nonumber \\
\Gamma_{3k-4} &= \hspace{-0.2cm}\raisebox{-0.65cm}{
$\stackrel{
\raisebox{-0.3cm}{
\scalebox{0.35}{\includegraphics{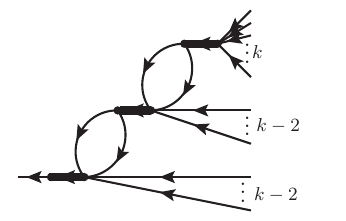}}
}}{\scriptscriptstyle {\cal O}(\lambda^{2/(k-1)})\quad} $
}\hspace{-0.3cm}
+
\raisebox{-0.75cm}{
$\stackrel{
\raisebox{-0.3cm}{
\scalebox{0.35}{\includegraphics{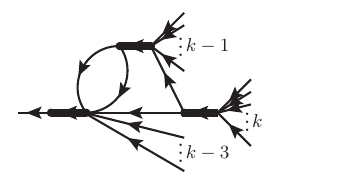}}
}}{\scriptscriptstyle {\cal O}(\lambda^{2/(k-1)})\quad} $
}\hspace{-0.3cm}
+
\raisebox{-0.85cm}{
$\stackrel{
\raisebox{-0.3cm}{
\scalebox{0.35}{\includegraphics{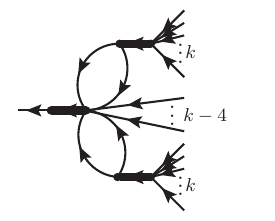}}
}}{\scriptscriptstyle {\cal O}(\lambda^{2/(k-1)})\quad} $
}\hspace{-0.3cm} 
 \nonumber
\end{align}
\caption{Vertices that set the perturbative leading-order (first row) and next-to-leading order (second and third row) correction to the mean-field result for $k>2$.}
\label{fig:4}
\end{figure}

The first perturbative correction to the equation of motion constructed in this way (starting with the vertex $\Gamma_k$) is set by the vertex $\Gamma_{2k-2}$, which includes the two-particle memory function and is shown in the first line of Fig.~\ref{fig:4}. It induces a correction to the mean-field result that is suppressed by ${\cal O}(\lambda^{1/(k-1)})$, which dominates over the ${\cal O}(\lambda)$ renormalization of the $\Gamma_k$ vertex. Next-to-leading order corrections are shown in the second and third line of Fig.~\ref{fig:4}.  They are set by a second-order diagram that involves the three-particle memory function as well as three third-order diagrams that describe more complicated two-particle correlations. Perturbatively, there are at least $k-1$ processes that are of lower order than the simple ${\cal O}(\lambda)$ mean-field correction discussed in the previous Sec.~\ref{sec:effactionA}.

 The leading perturbative correction to the mean-field result is thus of order $\delta n \sim {\cal O}(\lambda^0)$, which implies
\begin{align}
\delta n_{\rm pert.}(t) &= \frac{B_k}{(Dt)^{d/2}} .
\end{align}
The coefficient $B_k$ follows from a solution of 
\begin{align}
\partial_t \delta n &= - k^2 \lambda n_0^{k-1} \delta n - n_0^{k-2} \int_0^t dt' \, \Gamma_{2k-2}(t-t') n_0^k(t') 
\end{align}
with
\begin{align}
\Gamma_{2k-2}(s) &= \frac{k^3(k-1)^2}{2} \lambda^2 S_{2}(s) \sim s^{d/2-1} ,
\end{align}
where the symmetry factor accounts for two fusion vertices and the $k (k-1)$ ways of connecting the lines in the loop integral. In $d=1$ (which is the relevant dimension for the perturbative correction; cf. Fig.~\ref{fig:1}), we find $B_k = k^2\Gamma(\tfrac{k}{k-1})/(2\sqrt{2} (k+1) \Gamma(\tfrac{1}{2} + \tfrac{1}{k-1}))$, which evaluates to \mbox{$B_{k=4} = 0.62$} and \mbox{$B_{k=5} = 1.60$}.

\begin{figure}[t!]
\begin{align}
\Gamma_{3} \Bigr|_{k=2} &= \raisebox{-0.45cm}{
$\stackrel{
\raisebox{-0.3cm}{
\scalebox{0.35}{\includegraphics{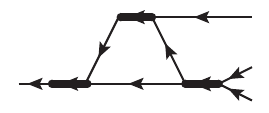}}
}}{\scriptscriptstyle {\cal O}(\lambda)} $
}\hspace{-0.3cm}
\nonumber
\end{align}
\caption{Leading-order correction to the mean-field result in the special case $k=2$. This correction is of the same order as the dimer memory correction shown in Fig.~\ref{fig:4}.}
\label{fig:5}
\end{figure}

Note that the above power counting for higher-order vertices does not apply for $k=2$ (where $d_c=2$). An example of a leading-order correction is the vertex $\Gamma_3$, which is of order ${\cal O}(\lambda^3)$ and shown in Fig.~\ref{fig:5} [note that a hypothetical ${\cal O}(\lambda^2)$ diagram similar to Fig.~\ref{fig:4} with one internal line is not 1PI]. In $d=3$, this vertex evaluates to~\cite{braaten02}
\begin{align}
\Gamma_{3} \Bigr|_{k=2} &= \frac{\lambda^3}{3 D \sqrt{D s}} + {\cal O}(\lambda^4)
\end{align}
and its contribution to the equation of motion will be suppressed by ${\cal O}(\lambda)$ compared to the mean-field term. The vertex $\Gamma_4$ induces a correction of the same order. Unlike for $k>2$, they are of the same order as the correction to the mean-field decay rate obtained by expanding the vertex $\Gamma_2$. Note that a similar mixing of different contributions in the effective action approach was noted by~\textcite{lee94}, where taking into account the vertex correction alone for \mbox{$k=2$} leads to a decay amplitude below $d_c$ at variance with renormalization group calculations. In the following, we focus on the case $k>2$, where the power counting is set by higher-order vertices.

\subsection{Nonperturbative corrections}\label{sec:effectiveactionC}

The perturbative results discussed in the previous section apply if the memory functions $S_2, S_3, \ldots, S_{k-1}$ that appear in the vertices $\Gamma_{2k-2}, \Gamma_{2k-3}, \ldots$ are finite. While the $k$-particle memory function $S_k$ is always finite above the critical dimension, this is only true for other memory functions if $d_c < d < 2/(k-2)$. In higher dimensions, some (or indeed all for $d\geq 2$) of them may contain logarithmic or power-law divergences, starting at $d=2/(k-2)$ with a logarithmic divergence in the vertex $\Gamma_{k+1}$; cf. Fig.~\ref{fig:5}. Such a cutoff dependence can have at least three different implications for scaling. (a) It can remain explicitly. (b) If a strong cutoff dependence can be removed by further renormalization, the scaling corrections depend on other parameters in addition to $\lambda$. (c) If the divergence is only superficial, summing the vertex to all orders will give a manifestly finite result.

Our calculations indicate that the latter case applies, i.e., the vertices summed to all order are finite and only depend on $\lambda$. Since in this case the only time dependence is introduced by the external fields in the effective action, the leading nonperturbative correction to mean-field scaling is set by the vertex $\Gamma_{k+1}$, which (by dimensional analysis) scales as ${\cal O}(\lambda^{1 + d/(d(k-1)-2)})$. This implies
\begin{align}
\delta n_{\rm nonpert.}(t) &= \biggl(\frac{\lambda}{D}\biggr)^{d_c d/2(d-d_c)} \frac{B_k}{(\lambda t)^{2/(k-1)}} \label{eq:nnonpert}
\end{align}
with a numerical coefficient $B_k$ that will be determined in the following. Provided that $d>4/(k-1)$, the nonperturbative contribution of the vertex $\Gamma_{k+1}$ to the equation of motion dominates over the perturbative correction discussed in the previous section, which is indicated by the dark blue shaded region in Fig.~\ref{fig:1}.

\begin{figure}[t!]
\subfigure{}{\begin{align}
\Gamma_{k+1} &= \raisebox{-0.35cm}{\scalebox{0.4}{\includegraphics{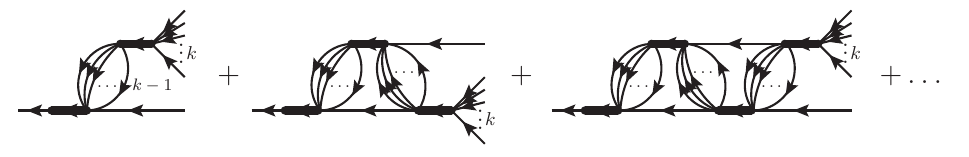}}}  \nonumber \\
&= \raisebox{-0.27cm}{\scalebox{0.4}{\includegraphics{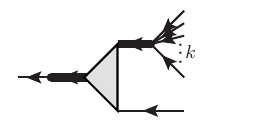}}} \nonumber
\end{align}}{(a)}  \\[1ex]
\subfigure{}{\scalebox{0.6}{\includegraphics{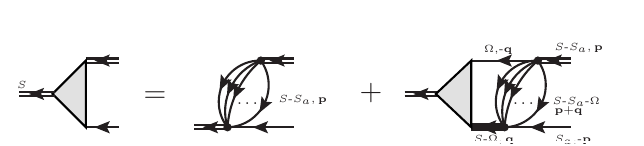}} \\}{(b)}
\caption{Bethe-Salpeter equation for the $(k+1)$-body vertex that contributes to the scaling correction above the critical dimension.}
\label{fig:6}
\end{figure}

\begin{figure*}[t!]
\scalebox{0.45}{\includegraphics{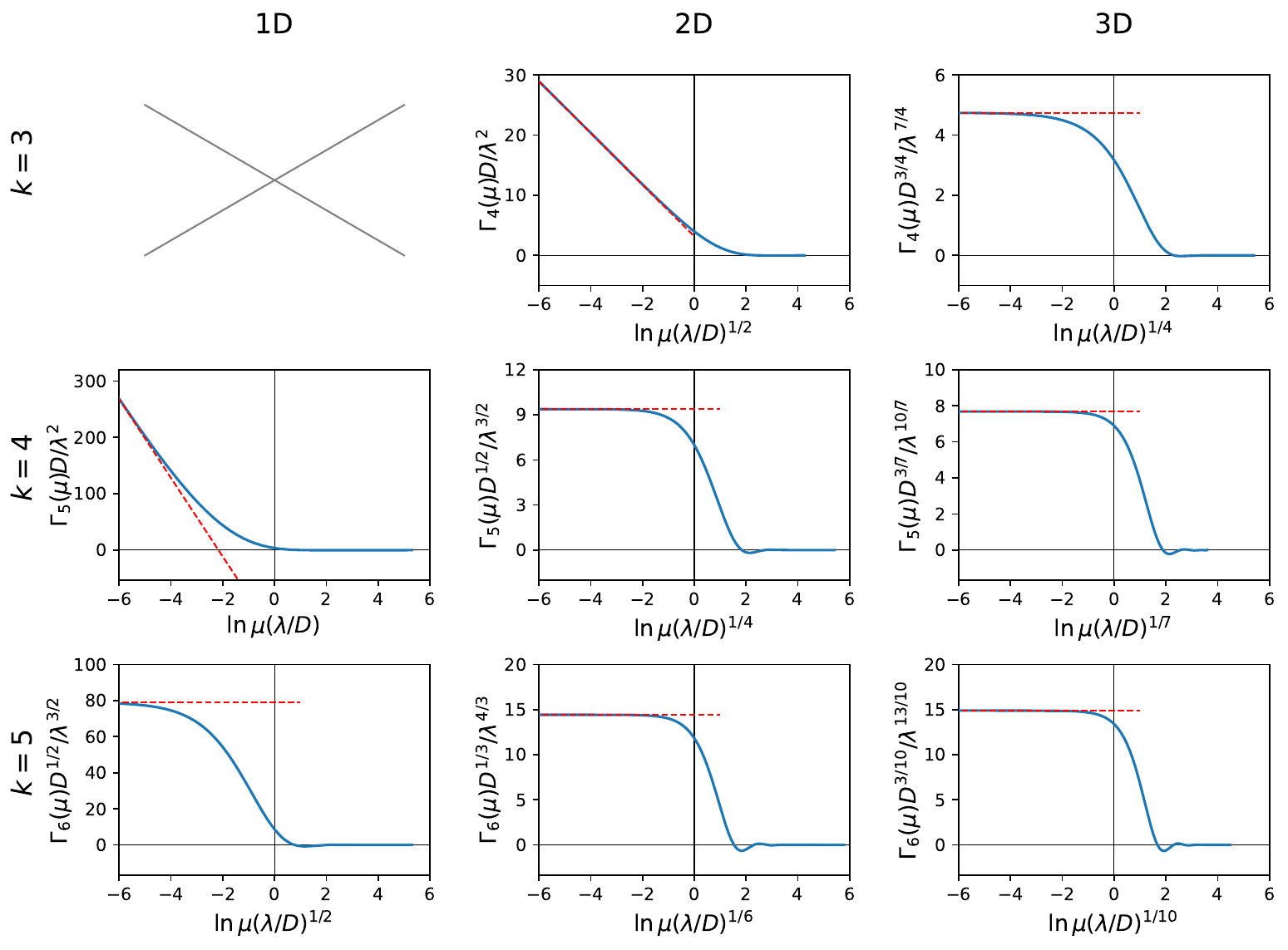}}
\caption{Dimensionless scaling function of the $(k+1)$-particle vertex $\Gamma_{k+1}$. The parameter choices correspond to the solid points in Fig.~\ref{fig:1} [we exclude $(k,d) = (3,1)$ as this is the marginal dimension of the process]. Blue lines indicate the full numerical result obtained from Eq.~\eqref{eq:bethe3b} and red dashed lines mark the small-$\lambda$ limit.}
\label{fig:7}
\end{figure*}

In general, it is not possible to sum a vertex with more than $k$ ingoing lines to all orders. To determine the vertex $\Gamma_{k+1}$, however, we apply methods developed for cold quantum gases to compute the three-body scattering matrix exactly~\cite{braaten02,bedaque99a,bedaque99b} (for a review, see Ref.~\cite{braaten06}). The first three terms that contribute to the vertex $\Gamma_{k+1}$ are shown in Fig.~\ref{fig:6}(a) (note that, for $k=2$, the first term is not 1PI and the vertex function starts with the second term). These diagrams are summed to all orders using a vertex that is implicitly defined as shown in Fig.~\ref{fig:6}(b). Unlike the $k$-particle vertex $\Gamma_k$, this is not a geometric series but represents an integral equation, which is given by
\begin{align}
&\Gamma_{k+1}(S | S_a, {\bf p}) = (-\lambda)^2 k! k^2  
S_{k-1}(S - S_a, {\bf p}) \nonumber \\*
&\quad + \int_{\rm BW} \frac{ds}{2\pi i} \int \frac{d^dq}{(2\pi)^d}\, 
\Gamma_{k+1}(S | s, {\bf q}) \frac{1}{s + D {\bf q}^2} (- \lambda) \nonumber \\*
&\qquad\qquad \times k^2 (k-1)! S_{k-1}(S-S_a-s, {\bf p} + {\bf q}) . \label{eq:integralequation}
\end{align}
Here, the vertex is a function of a total frequency $S$ and the frequency of the ingoing particle line $S_a$, as well as a relative momentum ${\bf p}$ between the ingoing particle and the $k$-particle line. The inhomogeneous term in the integral equation~\eqref{eq:integralequation} corresponds to the first diagram in Fig.~\ref{fig:6}(a) or~\ref{fig:6}(b), where $k!$ is a symmetry factor for the different ways of combining the internal loop lines and an additional factor of $k^2$ stems from the two fusion vertices; cf. Fig.~\ref{fig:2}. The homogenous term of the integral equation corresponds to the second term in Fig.~\ref{fig:6}(b). It involves the vertex function with loop frequency $s$ and momentum ${\bf q}$ and a single-particle propagator, as well as the $k$-particle propagator, which as before is replaced by its mean-field value $-\lambda$. In addition, the integrand contains the memory function for $k-1$ particles as a subdiagram, where $k^2 (k-1)!$ is now a symmetry factor that accounts for the different ways of combining the lines in the loop and the ingoing and outgoing line.

The frequency integration is evaluated using the residue theorem, which picks up the pole at $s = - D q^2$. To determine the running of the vertex at a small momentum or frequency scale $\mu$, we set the ingoing particle frequency equal to its value at the diffusion pole, $S_a = - D p^2$, as well as $S=0$, such that $\mu = |{\bf p}|$~\cite{braaten02}. This gives
\begin{align}
&\Gamma_{k+1}(
{\bf p}) = (-\lambda)^2 k! k^2 S_{k-1}(
D p^2, {\bf p}) \nonumber \\*
&\quad + \int \frac{d^dq}{(2\pi)^d}\, \Gamma_{k+1}(
{\bf q}) (-\lambda) k^2 (k-1)! \nonumber \\*
&\qquad\qquad \times S_{k-1}(
D (p^2 + q^2), {\bf p} + {\bf q}) . \label{eq:bethe3b}
\end{align}
The loop-angle integral over $S_{k-1}$ is performed in closed analytical form. The resulting one-dimensional integral equation is a Fredholm equation of the second kind that is solved using numerical standard algorithms such as the Nystrom method~\cite{press02} (for an introduction to the method applied to the three-body problem in ultracold quantum gases, see~\cite{barth15}). The integral equation is solved taking into account a momentum range $q \in [0,\Lambda]$ while retaining the explicit (divergent) cutoff dependence in the integration kernel and the inhomogeneous term. 

Figure~\ref{fig:7} shows the result for $\Gamma_{k+1}(p=\mu)$ for a range of dimensions $d=1,2,3$ and parameters $k=3,4,5$, where we exclude the case $(k,d) = (3,1)$ as this is the marginal dimension for this process. As is apparent from the figure, the vertex functions are finite and strongly suppressed at large $\mu$. We checked that the solution is independent of the cutoff scale and takes a scaling form that depends only on a dimensionless scaling variable $\mu (\lambda/D)^{d_c/2(d-d_c)}$. This confirms the power counting established at the beginning of the section. The perturbation solves
\begin{align}
\partial_t \delta n &= - k^2 \lambda n_0^{k-1} \delta n + \Gamma_{k+1}(\mu=0) n_0^{k+1}(t) ,
\end{align}
which reproduces the result~\eqref{eq:nnonpert} with 
$B_k = \Gamma_{k+1}(\mu=0)  (D/\lambda)^{d/(d(k-2)-2)}/(\lambda k(k-1) [k(k-1)]^{2/(k-1)})$
. The static limit \mbox{$\Gamma_{k+1}(\mu=0)$} is indicated in Fig.~\ref{fig:7} by the red dashed lines.

We conclude this section by discussing the special case where the integral equation~\eqref{eq:bethe3b} sums a logarithmic divergence of the memory function $S_{k-1}$. Such a logarithmic divergence occurs for $d = 2/(k-2)$, which can be seen directly  from Eq.~\eqref{eq:Gl} (changing \mbox{$k\to k-1$} to describe $S_{k-1}$). In integer dimensions, this corresponds to the two cases $(k,d)=(4,1)$ and $(k,d)=(3,2)$; cf. Fig.~\ref{fig:7}.

First, for $(k,d)=(4,1)$, we have for small $\lambda$:
\begin{align}
\Gamma_{5}(\mu) \bigr|_{k=4,d=1} &= - \frac{128 \sqrt{3} \lambda^2}{\pi D} \Bigl[\ln (\tfrac{\mu \lambda}{D}) + 2.24\Bigr] \label{eq:gamma5} .
\end{align}
However, as discussed above, for the parameter choice $(k,d)=(4,1)$, the vertex $\Gamma_{k+1=5}$ is subleading compared to the perturbative result (which is set instead by the vertex $\Gamma_{2k-2=6}$; cf. Sec.~\ref{sec:effectiveactionB}).

Second, for $(k,d)=(3,2)$, we have
\begin{align}
\Gamma_{4}(\mu) \bigr|_{k=3,d=2} &= - \frac{27 \lambda^2}{2\pi D} \Bigl[\ln (\mu \sqrt{\tfrac{\lambda}{D}}) - 0.73\Bigr] \label{eq:gamma4} .
\end{align}
This term sets the leading-order correction at $(k,d)=(3,2)$. Solving the equation of motion including this logarithmic correction gives instead of Eq.~\eqref{eq:nnonpert}
\begin{align}
n(t) \bigr|_{k=3,d=2} &= \frac{1}{(6\lambda t)^{1/2}} - \frac{3 \ln \tfrac{\lambda}{D^2 t} + 1.73}{8 \pi D t} + {\cal O}\biggl(\frac{\lambda^{1/2}}{t^{3/2}}\biggr) ,
\end{align}
which contains a logarithmic correction in time, too. Note that, beyond the leading-order correction, there can be additional corrections that include the range of the reaction potential~\cite{braaten02,tan08,mestrom19}.
 
 \section{Summary and outlook}\label{sec:summary}
 
In summary, we have discussed beyond-mean field corrections to the late-time dynamics of absorptive reaction-diffusion processes with $k$-particle annihilation. Using a Bose gas representation of the process, we link scaling corrections to few-boson scattering amplitudes, which capture memory effects of past reactions. Importantly, the leading corrections are not just given by a small renormalization of the $k$-particle reaction rate but by memory effects that involve a larger number of particles. This gives rise to two distinct regimes---a perturbative one and a nonperturbative one---with different scaling exponents for the corrections. The main results of this work are summarized in Fig.~\ref{fig:1}. 

For the specific case of absorptive reaction-diffusion processes, further work to compute correlation functions~\cite{lee94} or applications to fusion processes $k A \to l A $ with $l<k$~\cite{peliti86} and reactions involving multiple reactant species~\cite{lee95,howard96,konkoli99,konkoli00} appear straightforward. It is worth pointing out that, in evaluating higher-order corrections, we apply techniques that are well-known to describe few-particle scattering in ultracold quantum gases and nuclear physics, but that are perhaps not widely used in other fields. While this paper provides an application to a particular class of reaction-diffusion systems, it would be interesting to apply these methods more broadly.

\begin{acknowledgements}
I thank B. Mehlig and W. Zwerger for discussions and comments. This work is supported by Vetenskapsr\aa det (Grant No. 2020-04239).
\end{acknowledgements}

\bibliography{bib_reaction}

\end{document}